\documentclass[reprint,amsmath,amssymb,aps,superscriptaddress]{revtex4-2}

\usepackage[retainorgcmds]{IEEEtrantools}
\usepackage{float}
\usepackage{amsmath, bm,mathtools} %mathtools give \Aboxed
\usepackage{amssymb}
\usepackage{subcaption}
\usepackage{dsfont}
\usepackage{hyperref}
\usepackage{dsfont}
\usepackage{slashed} %Feynman notation
\usepackage{mathtools} %commands like xrightarrow
\usepackage[dvipsnames]{xcolor}

\usepackage{tikz}
\usepackage{tkz-euclide}
\usetikzlibrary{decorations.pathmorphing}	% To 
\usetikzlibrary{shapes}	% To draw a cylinder
\tikzset{every picture/.style={line width=1}}

\newcounter{qnumber}

\definecolor{c1}{rgb}{0.121569, 0.466667, 0.705882}
\definecolor{c2}{rgb}{1., 0.498039, 0.054902}
\definecolor{c3}{rgb}{0.172549, 0.627451, 0.172549}
\definecolor{c4}{rgb}{0.839216, 0.152941, 0.156863}
\definecolor{c5}{rgb}{0.580392, 0.403922, 0.741176}
\definecolor{c6}{rgb}{0.54902, 0.337255, 0.294118}
\definecolor{c7}{rgb}{0.890196, 0.466667, 0.760784}
\definecolor{c8}{rgb}{0.498039, 0.498039, 0.498039}
\definecolor{c9}{rgb}{0.737255, 0.741176, 0.133333}
\definecolor{c10}{rgb}{0.0901961, 0.745098, 0.811765}

\usepackage{color}

\usepackage{graphicx}% Include figure files
\usepackage{dcolumn}% Align table columns on decimal point
\usepackage{bm}% bold math
\usepackage{placeins} %force figure

\begin{document}

\preprint{APS/123-QED}

\title{Enhancing Constraints on Ultralight Axion Dark Matter from Gravitational Capture}

\author{Pierce Giffin}
\email{pgiffin@princeton.edu}
\affiliation{Department of Physics and Santa Cruz Institute for Particle Physics,
University of California Santa Cruz, Santa Cruz, CA 95064, USA}
\affiliation{Department of Physics, Princeton University, Princeton, NJ 08544, USA}

\author{Pankaj Munbodh}
\email{pmunbodh@anl.gov}
\affiliation{High Energy Physics Division, Argonne National Laboratory, Lemont, IL 60439, USA}
\affiliation{Department of Physics, Grainger College of Engineering,
University of Illinois Urbana-Champaign, Urbana, IL 61801, USA}
\affiliation{Department of Physics and Santa Cruz Institute for Particle Physics,
University of California Santa Cruz, Santa Cruz, CA 95064, USA}

\author{Elisa G. M. Ferreira}
\email{elisa.ferreira@ipmu.jp}
\affiliation{Kavli IPMU (WPI), UTIAS, The University of Tokyo,
5-1-5 Kashiwanoha, Kashiwa, Chiba 277-8583, Japan
}
% \affiliation{Center for Data-Driven Discovery, Kavli IPMU (WPI), UTIAS,
% The University of Tokyo, Kashiwa, Chiba 277-8583, Japan}

%\date{\today}% It is always \today, today,
             %  but any date may be explicitly specified

\begin{abstract}
Ultralight axions can be gravitationally captured by massive bodies such as the Sun, producing solar-bound gravitational atom states that amplify the local dark matter density through Bose-enhanced capture. For axion masses in the range 
$10^{-14}\,\text{eV}\lesssim m_a \lesssim 10^{-13}\,\text{eV}$, this mechanism becomes exponentially efficient. We show that, for a representative decay constant $f_a\sim 3.5\times10^7$ GeV, the local axion dark matter density at Earth can grow to more than ten times the standard Galactic value of $\rho_0 = 0.4 \text{ GeV/cm}^{3}$. Incorporating this overdensity, we derive updated limits on the axion–photon coupling from existing satellite and terrestrial measurements and present improved projections for upcoming experiments.
\end{abstract}

%\keywords{Suggested keywords}%Use showkeys class option if keyword
                              %display desired
\maketitle

%\tableofcontents

\section{\label{sec:level1} Introduction}

One of the most compelling solutions to the strong CP problem is the QCD axion~\cite{PhysRevLett.38.1440, PhysRevD.16.1791, PhysRevLett.40.223, PhysRevLett.40.279}. Axion-like particles (ALPs) arise generically in extensions of the Standard Model and in string theory compactifications~\cite{Svrcek:2006yi, Green:1987mn, Witten:1984dg}, where they can produce a whole `axiverse' of ultralight axions from topological considerations~\cite{Arvanitaki:2009fg}. These ultralight particles are well-motivated dark matter (DM) candidates. This candidate, also known as ultralight dark matter or axion DM, can describe all of the DM for masses in the range $10^{-24}\,\mathrm{eV} \lesssim m_a \lesssim 1 \,\mathrm{eV}$, or a fraction of it for lighter masses (for reviews, see~\cite{Hui:2021tkt, Ferreira:2020fam, Eberhardt:2025caq}).

Given the small mass and correspondingly large de Broglie wavelength of these bosonic DM candidates, gravitational focusing can occur when self-interacting bosonic particles in external gravitational potentials are trapped into bound states, leading to an exponential growth in the DM density~\cite{Budker:2023sex}. These clouds of ALP DM are also called halos\footnote{Although we call them halos, they are not virialized but consist of gravitationally bound particles.}, or, in the case of the Sun, solar halos. Such solar halos can enhance the local DM density by orders of magnitude relative to the fiducial value $\rho_{0} \simeq 0.4 \,\mathrm{GeV\,cm^{-3}}$, and may dominate over the background value inferred from Galactic dynamics alone. The phenomenological ramifications of solar and terrestrial axion halos was studied in Ref.~\cite{Banerjee_2020}, which remained agnostic about the halo formation mechanism. In this work, we employ the stimulated-capture mechanism detailed in Ref.~\cite{Budker:2023sex} to derive concrete updates to existing and projected bounds on the axion-photon coupling. Below we outline the relevant aspects of the mechanism.

The evolution of the gravitational atom is governed by the parameter 
\begin{equation}
\xi_{\rm foc} \equiv \frac{\lambda_{\rm dB}}{R_\star},
\end{equation}
where $\lambda_{\rm dB}$ is the de Broglie wavelength of the DM and $R_\star = 1/(GMm^2)$ is the characteristic length scale of the gravitational atom (much akin to the Bohr radius of the hydrogen atom). Here, $M$ is the mass of the body capturing the DM and $m$ is the mass of a DM particle.
Initially, the growth of the bound DM mass is linear in time. %and independent of $\xi_{\rm foc}$. 
This growth saturates to a small constant overdensity on a timescale $\propto \xi_{\rm foc}^{-4}$ when $\xi_{\rm foc} \lesssim 1$, for example, if we consider the Earth to be the source of the gravitational potential instead of the Sun. In this case, the waves are too fast to feel the gravitational potential strongly, leading to extremely small enhancements. 

On the other hand, for the regime that we are interested in ($\xi_{\rm foc}\gtrsim 1$), the linear growth quickly turns exponential as a consequence of stimulated capture arising from Bose enhancement. A critical DM density can be defined when the self-interaction energy becomes large enough to contend with the gravitational potential energy. Two possibilities arise once the critical density is reached, depending on the nature of the self-interactions. For repulsive self-interactions, the density saturates. For attractive self-interactions, an instability sets in and the solar halo undergoes a Bosenova explosion, ejecting relativistic bosons. 

As discussed in Ref.~\cite{Budker:2023sex}, the formation of the gravitational atom should not be affected in the presence of couplings to the Standard Model (SM) (in addition to self-interactions) owing to the lack of Bose enhancement.

Given this large enhancement in the density from the solar halo, and focusing on cases where the halo encompasses the Earth, in this paper, we study the impact of gravitational focusing on current~\cite{Nishizawa:2025_1,Nishizawa:2025_2,10.1093/ptep/ptag097, Yamamoto:2020, Chandra, CAST:2024, Sulai_2023, arza2025searchultralightdarkmatter} and proposed~\cite{Twisted_Anyon_Cavity, ADBC, PhysRevLett.133.111003, DANCE,Oshima_2023, ABRACADABRA} experimental searches for axions sensitive to the axion-photon coupling in the mass range $10^{-14}\text{ eV} \lesssim m_a \lesssim 10^{-13}\text{ eV}$.

We will take as benchmark a periodic potential for the axion $a$ of the form
\begin{equation}
    V(a) \supset - m_a^2 f_a^2 \cos \left (\frac{a}{f_a} \right ),
\end{equation}
\begin{equation}
    V(a) \supset \frac{1}{2}m_a^2 a^2 + \frac{\lambda}{4!} a^4,
\end{equation}
where $\lambda = -m_a^2/f_a^2$ is a dimensionless coupling that characterizes the self-interactions of the axion field, and $m_a$ is identified as the mass of the axion. Since $\lambda<0$, the potential is attractive. We further assume that the axion couples to the Standard Model (SM) through the axion-photon coupling
\begin{equation}
    V \supset -\frac{1}{4}g_{a\gamma\gamma}a F_{\mu\nu}\tilde{F}^{\mu\nu}.
\end{equation}
Though the  KSVZ and DFSZ QCD axion models fix $g_{a\gamma\gamma} \sim \frac{\alpha}{2\pi}\frac{1}{f_a}$ and $m_af_a\sim m_\pi f_\pi$ (where $\alpha$ is the fine-structure constant and $f_\pi$ and $m_\pi$ are the pion decay constant and mass, respectively), we consider here generic models of axion-like particles and treat $m_a$, $f_a$, and $g_{a\gamma\gamma}$ as independent parameters.

The paper is structured as follows. In Section~\ref{sec:evolution}, we discuss the evolution of the DM density through the formation of the gravitational atom. In Section~\ref{sec:overdensity}, we discuss the conditions on the $(m_a, f_a)$ parameters where a large dark matter overdensity is expected. For completeness, we also mention the effect of the overdensity in the case of a repulsive potential $\lambda = m_a^2/f_a^2>0$ of a light scalar. 
% For consistency, we keep the same notation throughout. 
In Section~\ref{sec:new_constraints}, we show the increase in sensitivity to the axion-photon coupling of current and future experiments. Finally, we summarize our main findings in Section~\ref{sec:conclusion}.

\section{Evolution during gravitational focusing}
\label{sec:evolution}
To track the evolution of the dark matter density at Earth's location, we will use the equations provided in Ref.~\cite{Budker:2023sex}. The treatment only considers the population of the ground state of the gravitational atom, neglecting contributions from higher excited states that give rise to nonlinearities. In fact, DM particles that scatter into bound higher excited states are expected to decay to the ground state, amplifying the enhancement further. Thus, our results in Section~\ref{sec:new_constraints} should be taken as conservative.

The dark matter density at a radius $r$ is given by
\begin{equation}
    \rho(r,t) = \rho(r=0,t) e^{-2r/R_\star},
\end{equation}
where $r=0$ is at the center of the Sun. 
% At Earth's location, the radius is fixed to $r \simeq 1$AU. 
Since we are interested in the effect of this overdensity on detectors on Earth or in its vicinity, we focus on the regime where the halo extends to $r \simeq 1\,\mathrm{AU}$. For notational simplicity, denote $\rho_{\rm S}(t) \equiv \rho(r=0,t)$.

The relaxation time $\tau_{\rm rel}$ is defined to be the typical time taken for a DM particle to change its velocity by an order one factor due solely to self-interactions,
\begin{equation}
    \tau_{\rm rel} (m_a, f_a) = \frac{64 m_a^3 f_a^4 v^2_{\rm dm}}{\rho^2_{0}}.
\end{equation}
In the case of gravitational focusing $\xi_{\rm foc}\gtrsim 1$, the growth is linear until a time $t_{\rm lin} \simeq 0.3 \tau_{\rm rel}$. For $ t_{\rm lin} < t < t_{\rm crit}$, the growth rate becomes exponential until the critical density is reached at $t=t_{\rm crit}$. The critical density $\rho_{\rm crit}$ is approximately given by
\begin{equation}
    \rho_{\rm crit} (m_a,f_a) \simeq 16 (GMm_a)^2 m_a^2 f_a^2.
\end{equation}
Ref.~\cite{Budker:2023sex} has verified that this heuristic expression provides an excellent approximation to the critical density as seen in numerical simulations.

For an attractive potential, which is the case for our benchmark axion model, a Bosenova explosion occurs, releasing a large fraction of the captured DM as relativistic axions when $t \simeq t_{\rm crit}$~\cite{Levkov_2017,Eby_2016}. These Bosenova events have been observed experimentally in condensed matter systems~\cite{Donley_2001} and are actively being searched for in astrophysical settings ~\cite{Bosenova_exp}. Dedicated studies on the precise amount of DM mass ejected in a solar halo Bosenova have yet to be performed. Hence, we assume, similarly to Ref.~\cite{Budker:2023sex}, that all the mass in the solar halo is immediately lost in the Bosenova explosion and that the formation of the gravitational atom restarts.

In summary, the evolution of the density $\rho_S(t)$ around the center of the Sun is given by
\begin{equation}
\label{eq:dsigma_dp_signal}
    \rho_S(t) =  \begin{cases} \rho_{\rm lin} \frac{t}{t_{\rm lin}} & \text{for} ~ t < t_{\rm lin}~, \\
    \rho_{\rm lin} e^{(t-t_{\rm lin})/t_{\rm lin}} & \text{for}~ t_{\rm lin} \lesssim t \lesssim t_{\rm crit}~,\\
    \rho_S(t \text{ mod } t_{\rm crit}) & \text{for} ~ t_{\rm crit} < t \text{  (attractive)}~,\\
    \rho_{\rm crit} & \text{for} ~ t_{\rm crit} < t \text{ (repulsive)}~,
    \end{cases}
\end{equation}
where $\rho_{\rm lin}$ denotes the density at the end of the linear growth phase and is given by
\begin{equation}
    \rho_{\rm lin}(m_a) = \pi^{-5/2} \left (\frac{2\pi}{v_{\rm dm}}\right )^3 (GMm_a)^3 h [(GMm_a)^2/v^2_{\rm dm}]\rho_0,
\end{equation}
where $h(x)$ is well approximated by 
\begin{equation}
    h(x)\simeq 0.12 s(x)/(1-s(x))
\end{equation} 
and $s(x) \simeq e^{-x-2+\sqrt{4+2x}}$. The critical time is given by
\begin{equation}
    t_{\rm crit}(m_a,f_a) = t_{\rm lin} \left [\ln\left (\frac{\rho_{\rm crit}}{\rho_{\rm lin}}\right ) +1 \right ].
\end{equation}

\begin{figure*}[tb]
    \centering
    \includegraphics[width=0.9\linewidth]{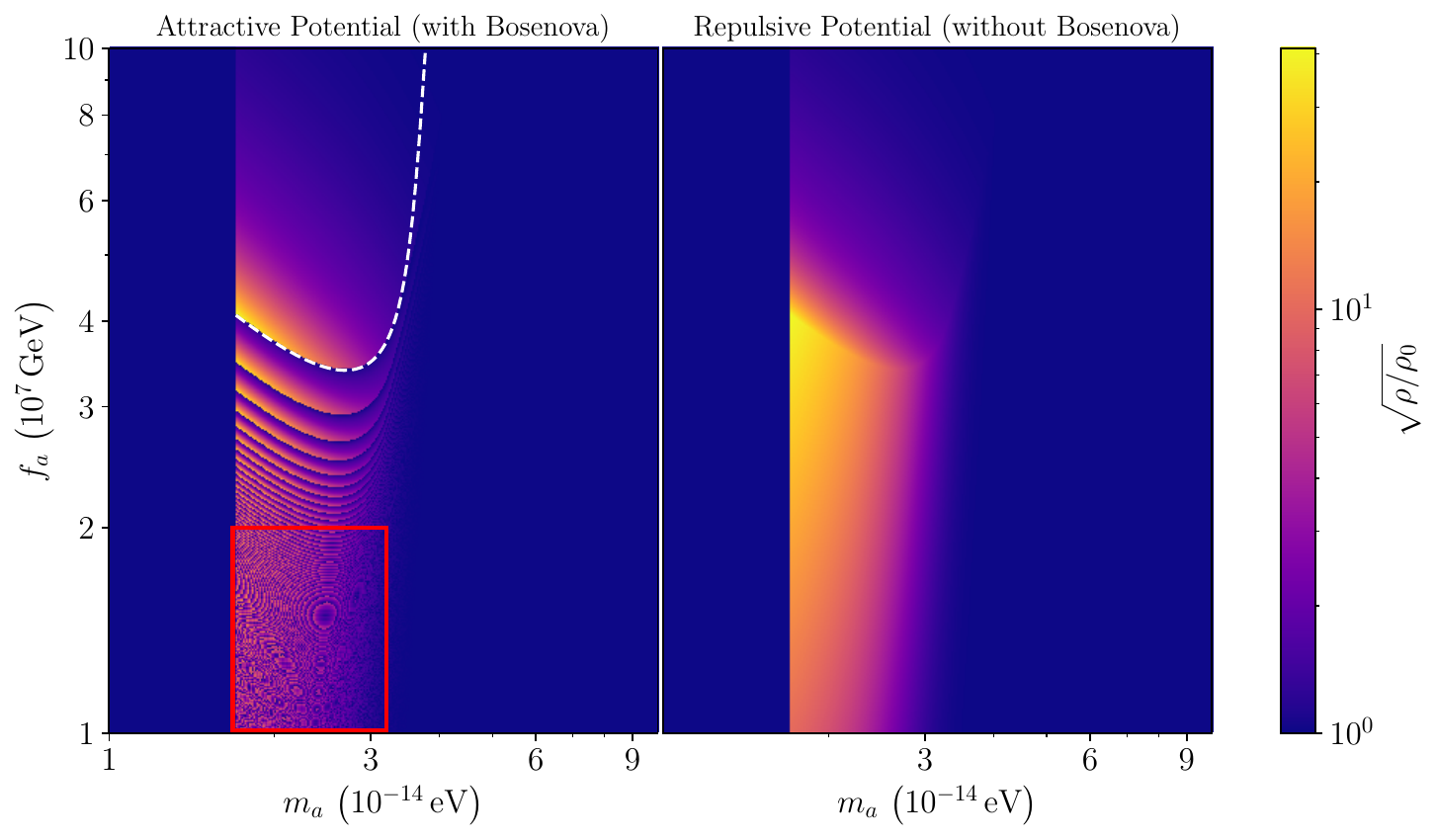}
    \caption{Dark matter overdensity factor ($\sqrt{\rho/\rho_0}$)  for the case of an attractive potential (left) and repulsive potential (right). Here we do not consider the scenario in which $\xi_{\rm foc}<1$ as the DM capture rate does not become exponential. The dashed white line corresponds to a choice of parameters such that the first Bosenova event occurs at 5 Gyr. Further edges for smaller values of $f_a$ indicate an increased number of Bosenova events occurring within the same timeframe. The region inside the red box corresponds to parameters that yield Bosenova events occurring more frequently than every $\sim 0.5$ Gyr. Hence, in this region, we cannot be certain as to which phase in the capture cycle we currently reside and can no longer trust our results.}
    \label{fig:Heatmap}
\end{figure*}

\section{DM overdensity from self-interactions}
\label{sec:overdensity}
%In this section, we identify the parameter ranges of interest in $(m_a,f_a)$ for gravitational focusing. %and discuss the experiments that benefit from improved sensitivity.

The region where gravitational focusing is relevant ($\xi_{\rm foc}\gtrsim 1$) corresponds to a lower bound of $m_a\gtrsim1.7\times 10^{-14}$ eV. Additionally, the calculation in Ref.~\cite{Budker:2023sex} always treats the potential as $\sim1/r$, implying that it is only valid if the radius of the ground state exceeds the solar radius $R_\star > R_{\rm sun}$, translating to a lower bound of $m_a \lesssim 2\times 10^{-13}$ eV. The enhancement is only exponential in the range $10^7\, \text{GeV} \lesssim f_a \lesssim 10^8\,\text{GeV}$. At larger values of $f_a$, the relaxation time increases, extending the linear phase and as a result preventing the onset of the exponential phase. At lower values of $f_a$, the critical density is low enough such that we do not have substantial enhancement.
% we do not have a substantial enhancement.

In Fig.~\ref{fig:Heatmap}, we show the dark matter overdensity ($\sqrt{\rho/\rho_0}$) today (at $t\simeq 5$ Gyr) in the plane of $(m_a, f_a)$ for an attractive potential (left) and for a repulsive potential (right). Although our benchmark axion model has an attractive potential, we also show the repulsive case where the density saturates to (at least) the critical density, to highlight the difference between the two cases, namely the non-occurrence of Bosenova collapse~\cite{Dmitriev:2021utv}. For context, Ref.~\cite{Fan:2016rda} discusses the phenomenology of an ultralight dark matter candidate with repulsive self-interactions, including the model-building challenges encountered. The dashed white line on the left panel gives the contour on the $(m_a,f_a)$ plane that leads to one Bosenova at $t\simeq 5$ Gyr and for which the critical density is reached today. For smaller values of $f_a$, the critical density decreases and multiple Bosenovas may occur in a 5 Gyr timescale. This feature can be clearly seen from the ridges below the white contour in Fig.~\ref{fig:Heatmap}, which get closer together as $f_a$ is lowered.

We apply a cutoff in Fig.~\ref{fig:Heatmap} at $m_a\sim1.7\times10^{-14}$ eV as this is when $\xi_{\rm foc}\sim1$, signifying that the capture mechanism is no longer able to grow exponentially and only grows linearly for smaller masses. We additionally enclose a region in Fig.~\ref{fig:Heatmap} with a red box corresponding to Bosenova events occurring more frequently than $0.5$ Gyr. As $f_a$ decreases, the frequency of Bosenova events increases, and the precise details on the age of the solar system and the resetting of the capture mechanism become relevant to determining the local dark matter density today. We caution that the results are not to be trusted in this region.

\begin{figure*}[t!]
    \centering
    \includegraphics[width=0.85\linewidth]{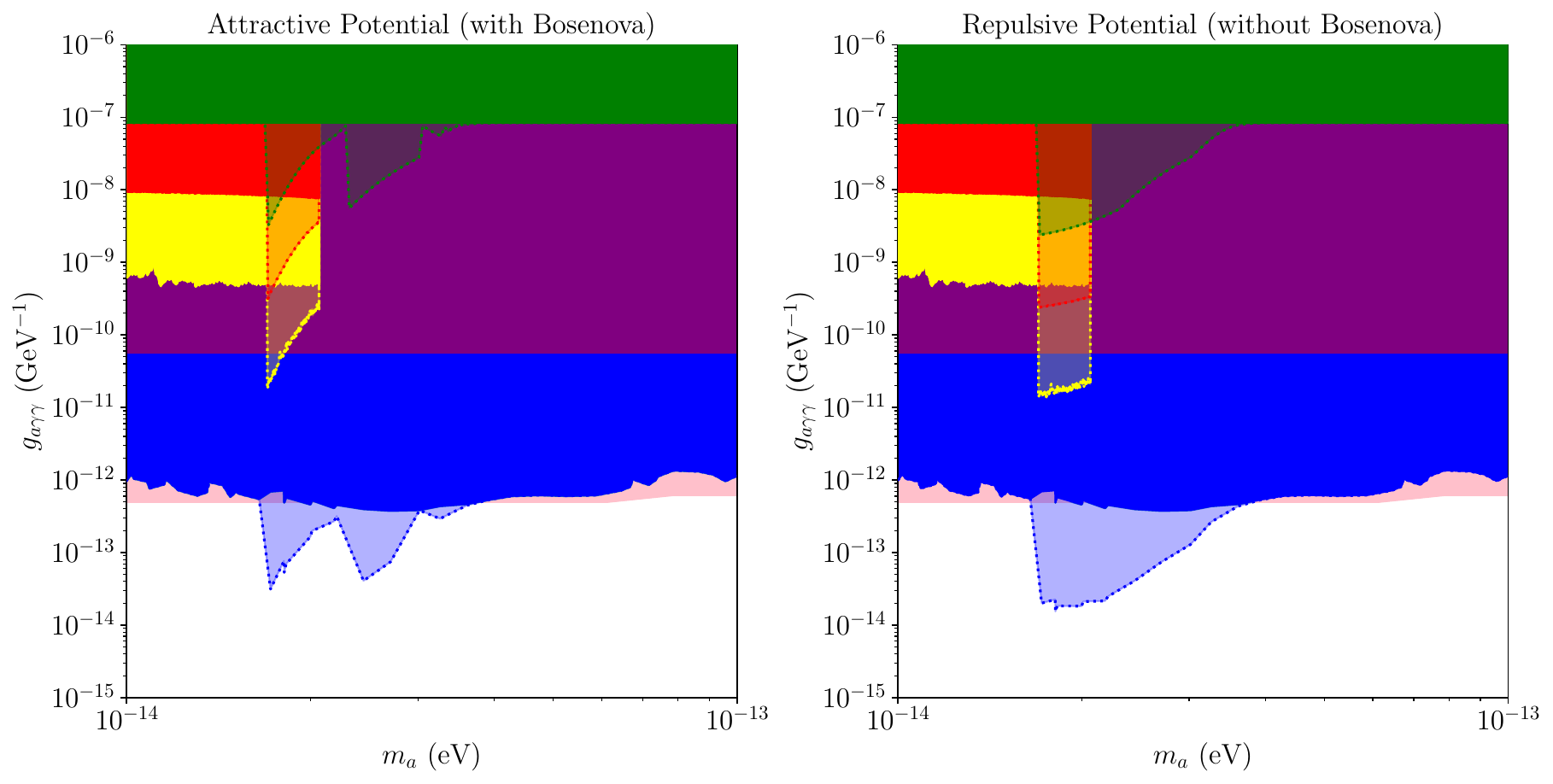}
    \caption{Updated constraints for the axion-photon coupling, $g_{a\gamma\gamma}$, as a function of the axion mass, $m_a$, for the case of an attractive potential (left) and repulsive potential (right). In both figures, $f_a=3.5\times10^7\,\text{GeV}$ was used as a benchmark scenario. The current \textit{Suzaku} satellite~\cite{Yamamoto:2020} (green), SNIPE-Hunt experiment~\cite{Sulai_2023} (red), GPEX experiment~\cite{arza2025searchultralightdarkmatter} (yellow),  and Eskdalemuir observatory~\cite{Nishizawa:2025_2} (blue) constraints are shown in the solid shaded regions alongside the Chandra~\cite{Chandra} (pink) and CAST~\cite{CAST:2024}(purple) limits. Enhanced constraints from this work are shown with dotted lines with light shading. The sharp transition at $m_a\sim 1.7\times10^{-14}$ eV is an artifact of the cut-off imposed at $\xi_{\rm foc} \sim 1$.}
    \label{fig:Constraints}
\end{figure*}

If the DM overdensity in the solar system is large enough, it may cause planets and asteroids to measurably deviate from their predicted orbits~\cite{Pitjev:2013sfa, Tsai:2022jnv}. This constrains the overdensity to lie below $\rho \lesssim 10^5\rho_0$ at $r=1$ AU. From Fig.~\ref{fig:Heatmap} in the $m_a$ mass range that we consider, we remain well below these constraints. A number of methods have been proposed in the literature to tighten the bounds on the local DM density near the Sun and the planets, including the Moon~\cite{Liang:2025bqr, Budker:2024bzj, Tsai:2025wuf}, potentially probing overdensities in the range considered here.

Additionally, measurements of the matter power spectrum at large scales constrain the strength of DM self-interactions, as this would impact structure formation~\cite{Arvanitaki:2014faa, Fan:2016rda, Cembranos:2018ulm}. For $10^{-14} \lesssim m_a \lesssim 2\times 10^{-13}$ eV, this would constrain the axion decay constant to $f_a\gtrsim 2-30\times 10^5$ GeV, which is always satisfied in Fig.~\ref{fig:Heatmap}. We remark that the effect of self-interactions on perturbations during radiation domination may result in a stronger bound on $f_a$. Although no dedicated study has been performed, Ref.~\cite{Budker:2023sex} provides a crude estimate of the bound obtained from the matter power spectrum.

Finally, let us comment briefly on some of the viable production mechanisms.
For the range of $m_a$ that we consider, the values of $f_a$ in Fig.~\ref{fig:Heatmap} are much too small for the axions to be produced from the standard misalignment mechanism whilst satisfying the relic density constraint. However, there are several variants of the misalignment mechanism that allow the axion to be all of the dark matter for the values of $f_a$ shown in Fig.~\ref{fig:Heatmap}. The kinetic misalignment mechanism~\cite{Eroncel:2022vjg, Eroncel:2024rpe}, where the axion is given a large initial velocity, provides a natural framework to achieve this. Alternatively, one can consider the large misalignment mechanism (at the cost of severe fine-tuning)~\cite{Arvanitaki:2019rax} or the bubble misalignment mechanism (at the cost of allowing for a first-order phase transition arising from hidden $SU(N)$ gauge sectors)~\cite{Lee:2024oaz}. 

%\PM{To look into coherence time. Add comment to say that we don't care about coherence time because we are always in exponential regime.}
%\FloatBarrier

We present here the region of $(m_a,\,f_a)$ parameter space that produces the largest possible enhancement to illustrate the impact on direct detection bounds of ALP dark matter. 

%{Even for values of $f_a$ compatible with formation mechanisms such as the misalignment mechanism, an $\mathcal{O}(1\text{--}10)$ enhancement can still be obtained.}\PM{Last sentence does not apply.}

\begin{figure*}[t!]
    \centering
    \includegraphics[width=0.9\linewidth]{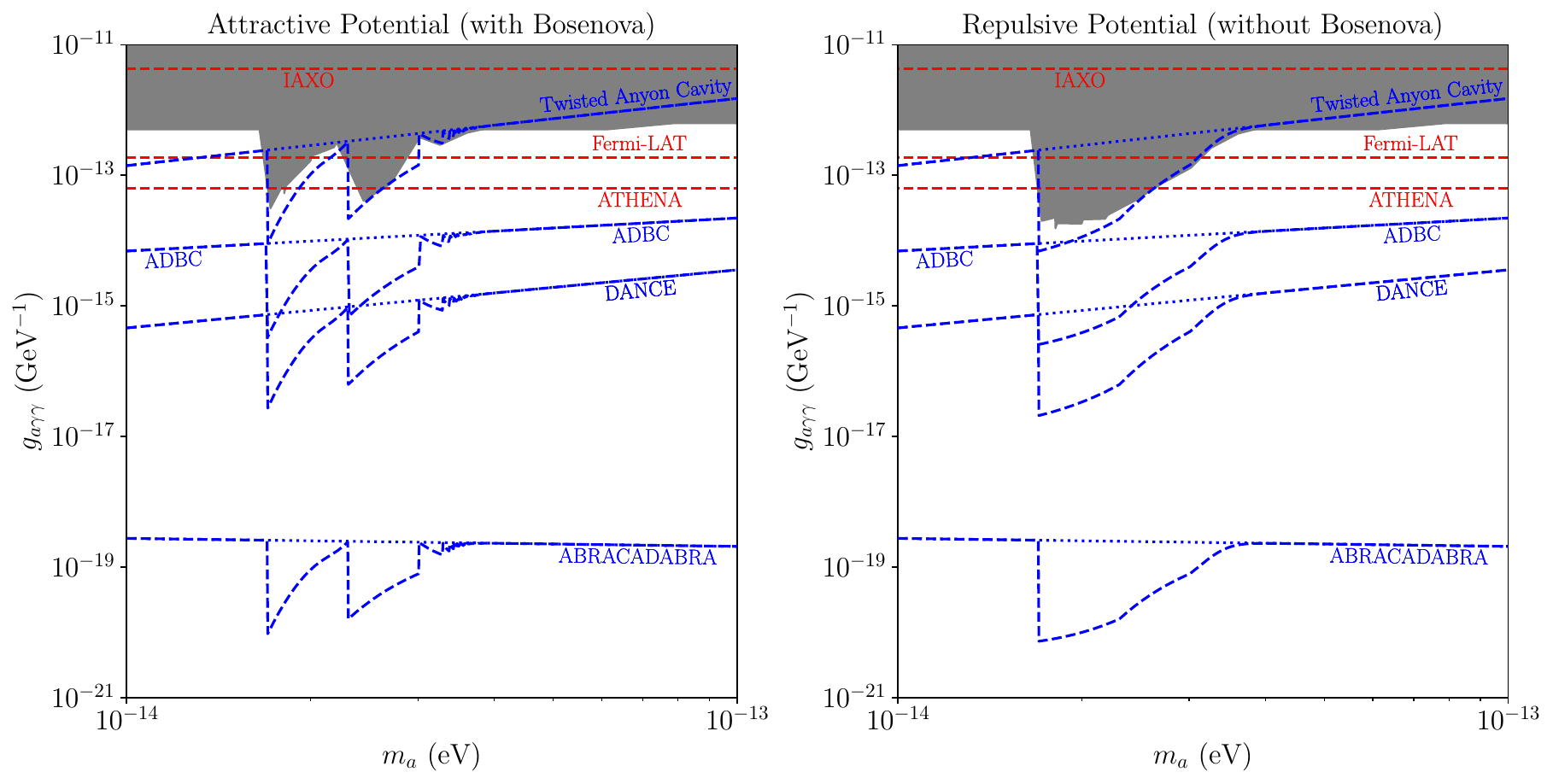}
    \caption{Projections for the axion-photon coupling, $g_{a\gamma}$, as a function of the axion mass, $m_a$, for the case of an attractive potential (left) and repulsive potential (right). In both figures, $f_a=3.5\times10^7\,\text{GeV}$ was used as a benchmark scenario. Experiments sensitive to the local dark matter density (Twisted Anyon Cavity~\cite{Twisted_Anyon_Cavity}, ADBC~\cite{ADBC, PhysRevLett.133.111003}, DANCE~\cite{DANCE,Oshima_2023}, 
    %SRF\cite{Heterodyne}, PG: Don't think we need to include this one
    and ABRACADABRA ~\cite{ABRACADABRA}) are shown as blue lines. The dotted lines show the unaltered constraints, while the dashed lines show improved constraints from an enhancement of the local DM density. Other relevant experiments that are insensitive to the local dark matter density (IAXO~\cite{IAXO}, Fermi-LAT~\cite{Fermi_LAT}, and ATHENA~\cite{ATHENA}) are shown in dashed red lines. Enhanced constraints from Fig.~\ref{fig:Constraints} are shown in the solid gray region. The sharp transition at $m_a\sim 1.7\times10^{-14}$ eV is an artefact of the cut-off imposed at $\xi_{\rm foc} \sim 1$.}
    \label{fig:Projections}
\end{figure*}

\section{Updated constraints on axion-photon coupling $g_{a\gamma\gamma}$}
\label{sec:new_constraints}

% Due to the degeneracy in $g_{a\gamma\gamma}\sqrt{\rho_0}$ that many axion dark matter direct detection experiments face, the relevant constraints placed can be improved by a factor of $\sqrt{\rho/\rho_0}$ if the local dark matter density is enhanced by gravitational focusing.

The improvement in sensitivity arises because many direct detection experiments probe the amplitude of the axion field, which scales as $\sqrt{\rho}$, leading to a degeneracy between the axion-photon coupling and the local dark matter density. As a result, an enhancement in the local density directly translates into stronger bounds on $g_{a\gamma\gamma}$ by a factor of $\sqrt{\rho/\rho_0}$. This effect is particularly relevant for experiments that rely on local measurements of oscillating electromagnetic fields induced by the axion, such as magnetometer-based searches and resonant cavity experiments. In contrast, helioscope searches and astrophysical probes that do not depend on the local dark matter density remain unaffected by gravitational focusing.

As Fig. \ref{fig:Heatmap} shows, we expect such enhancement of the local DM density within the range of $10^{-14}\,\text{eV}\lesssim m_a\lesssim10^{-13}\,\text{eV}$. Within this mass range, constraints have been placed from studies of Earth's magnetic fields using data from the \textit{Suzaku} satellite \cite{Yamamoto:2020,Suzuku} as well as the Eskdalemuir observatory \cite{Nishizawa:2025_2,Nishizawa:2025_1,Eskdalemuir}. The magnitude of the improvement depends on the achievable overdensity, which in turn is controlled by the axion self-interactions through $f_a$. For the benchmark value $f_a=3.5\times10^7\,\mathrm{GeV}$,  we find that the local density at Earth can be enhanced by more than an order of magnitude over the standard Galactic value across a significant portion of the mass range $10^{-14}\,\mathrm{eV}\lesssim m_a\lesssim10^{-13}\,\mathrm{eV}$. This corresponds to a noticeable shift of the exclusion limits toward smaller values of $g_{a\gamma\gamma}$, as illustrated in Fig.~\ref{fig:Constraints}. This translates into improvements in the constraints exceeding one order of magnitude in this mass window.

Additionally, several proposed direct detection experiments~\cite{Twisted_Anyon_Cavity,ADBC,DANCE, Oshima_2023,ABRACADABRA,Heterodyne} are similarly sensitive to DM overdensities. In Fig.~\ref{fig:Projections}, we show the projected sensitivities for a range of proposed experiments. For those relying on the local axion field amplitude, the enhancement effectively shifts their reach downward in $g_{a\gamma\gamma}$, extending the parameter space that can be probed. The projected sensitivities show a comparable improvement, exceeding one order of magnitude in parts of this mass window. This demonstrates that solar gravitational capture can play an important role in strengthening the discovery potential of upcoming searches for ultralight axion dark matter.

\section{Conclusion}
\label{sec:conclusion}

We have studied the effect of gravitational focusing on the sensitivity of current and future experiments searching for a non-zero axion-photon coupling. 
We show that experiments could have an $\mathcal{O}(10)$ increase in sensitivity to $g_{a\gamma\gamma}$ in the mass range $10^{-14}\lesssim m_a \lesssim 10^{-13}$ eV due to the local DM overdensity.
This enhancement arises because many direct detection experiments probe the amplitude of the axion field, which scales as $\sqrt{\rho}$, so that an increase in the local density directly translates into stronger bounds on the axion-photon coupling.

In modeling the evolution of the gravitationally bound axion cloud, we have only considered the population of the ground state of the gravitational atom, neglecting contributions from higher excited states. Throughout this work, we have presented results corresponding to values of $f_a\sim10^7 - 10^8$ GeV that maximize the enhancement in order to illustrate the largest possible impact on direct detection bounds. However, particles populating excited states are expected to decay to the ground state, which would further amplify the density enhancement. Including these effects would therefore strengthen the resulting bounds. In addition, we have assumed that the Bosenova collapse is instantaneous once the critical density is reached and that all the dark matter mass is ejected, which may underestimate magnitude of the enhancement immediately following a Bosenova event. If a significant portion of the ALP DM remains after the Bosenova collapse, the sharp features in the enhanced constraint curves in Fig. ~\ref{fig:Constraints} and Fig. ~\ref{fig:Projections} would be smoothed out but will on average remain at similar orders of magnitude.

%Nevertheless, for values of $f_a$ compatible with standard formation mechanisms such as the misalignment mechanism, the enhancement, while more moderate, is still present. \PM{No enhancement for standard misalignment is present.} Even in this regime, the solar halo can still increase the local density by an order unity to order ten factor, which translates into a noticeable strengthening of the constraints on $g_{a\gamma\gamma}$.

These results demonstrate that gravitational capture in the Solar potential can play an important role in interpreting and improving searches for ultralight axion dark matter. The presence of a solar axion halo can enhance the discovery potential of both current experiments and future proposals sensitive to the local dark matter density. A more detailed treatment of the nonlinear evolution of the halo, including the dynamics of excited states and a refined description of the Bosenova instability, is left for future work.

\section*{Acknowledgments}
We thank M. Grant Roberts for helpful discussion. We also thank Wolfgang Altmannshofer and Pouya Asadi for useful comments on a draft of this manuscript. The research of P.G. and P.M. is supported in part by the U.S. Department of Energy grant number DE-SC0010107. The research of P.G. is supported in part by the Achievement Rewards for College Scientists Foundation (ARCS) 2025-2026 and by the U.S. Department of Energy grant number DE-SC0007968. The research of P.M. is supported in part by the UC Chancellor's Dissertation Year Fellowship. The work of PM at Argonne National Laboratory was supported by the U.S. Department of Energy
under contract DE-AC02-06CH11357. Kavli IPMU is supported by World Premier International Research Center Initiative (WPI), MEXT, Japan

\bibliography{bibliography}

\end{document}